\documentclass[showpacs,prl]{revtex4}%
\usepackage{graphicx}
\usepackage{dcolumn}
\usepackage{amsmath}
\usepackage{amsfonts}%
\usepackage{amssymb}%

\begin{document}
\title{Effect of phonon scattering by surface roughness on the universal thermal conductance}
\author{D.~H.~Santamore}
\author{M.~C.~Cross} \affiliation{Department of Physics, California
Institute of Technology 114-36, Pasadena, California 91125}
\date{\today   }

\begin{abstract}
The effect of phonon scattering by surface roughness on the thermal
conductance in mesoscopic systems at low temperatures is calculated using full
elasticity theory. The low frequency behavior of the scattering shows novel
power law dependences arising from the unusual properties of the elastic
modes. This leads to new predictions for the low temperature depression of the
thermal conductance below the ideal universal value. Comparison with the data
of Schwab et al. [Nature \textbf{404}, 974 (2000)] suggests that surface
roughness on a scale of the width of the thermal pathway is important in the experiment.

\end{abstract}
\maketitle

Thermal transport in mesoscopic systems at low temperatures shows
universal properties analogous to the quantized electrical
conductance \cite{L57}. Theoretical analyses of phonon transport
when the thermal wavelength becomes comparable to the dimensions
of the thermal pathway predict a thermal conductance $K$ that is
independent of many of the details of the geometry and material
properties \cite{ACR98}, and a universal value for $K/T=N_{0}\pi
^{2}k_{B}^{2}/3h$ at low enough temperatures \cite{RK98,B99}.
(Here $N_{0}$ is the number of modes with zero frequency at long
wavelengths, equal to $4$ for a single, free-standing elastic
beam.) These predictions have since been connected to more general
results on bounds on entropy transport at low temperatures
\cite{P83}, and to thermal transport by particles of arbitrary
statistics \cite{RK99}. The recent confirmation of the universal
thermal conductance in tiny silicon nitride devices \cite{SHWR00}
is an experimental tour de force. Although verifying the
predictions of a universal value of $K/T$ at low enough
temperatures, the experiments showed values of $K/T$ that
\emph{decrease} as temperature increases in the range of $0.084$K
to $0.2$K, before beginning to rise at higher temperatures as more
vibrational modes that can carry the heat are excited.

We \cite{SC00} and others \cite{SFYM99,KFFL99} have previously studied a
simplified treatment of this problem using a scalar model for the elastic
waves. However elastic waves in confined geometries have many unusual
features, such as modes with a quadratic dispersion relation $\omega\propto
k^{2}$ at small wave numbers $k$ and regions of anomalous dispersion
$d\omega/dk<0$, that are not captured in this simple model and might be
expected to have a strong influence on the low temperature transport.

In this letter we study the general problem of the scattering of phonons by
rough surfaces for a rectangular cross-section elastic beam using three
dimensional elasticity theory. The scattering of the low frequency modes that
contribute to the conductance at low temperatures depends on the detailed
properties of the elastic modes, and we find novel power law dependences for
the frequency dependence of the scattering off unstructured roughness that are
\emph{not} those anticipated by simple analogy with Rayleigh scattering. In
turn this low frequency behavior yields a depression of $K/T$ at low
temperatures with unexpected power laws, with a faster decrease of $K/T$ as
the temperature is raised than anticipated from the scalar model. Comparing
these predictions with the experiments of Schwab et al.\ \cite{SHWR00}
suggests that unstructured roughness is inadequate to explain the data:
instead we find a roughness length scale comparable to the width yields a fit
consistent with the experimental trends.

In the ballistic transport regime, the thermal conductance at temperature $T$
takes the form \cite{ACR98,RK98,B99}
\begin{equation}
K=\frac{\hslash^{2}}{k_{B}T^{2}}\sum_{m}\frac{1}{2\pi}\int_{\omega_{m}%
}^{\infty}\mathcal{T}_{m}(\omega)\frac{\omega^{2}e^{\beta\hslash\omega}%
}{(e^{\beta\hslash\omega}-1)^{2}}\,d\omega. \label{conductance}%
\end{equation}
where the sum is over the modes $m$ propagating in the structure and
$\omega_{m}$ is the cutoff frequency of the $m$th mode, and $\beta=1/(k_{B}%
T)$. The parameter $\mathcal{T}_{m}(\omega)$ is the transmission coefficient:
$\mathcal{T}_{m}(\omega)\neq1$ corresponds to a reduction in the transport due
to some scattering process. At low temperatures, only the modes with
$\omega_{m}=0$ contribute, and in the absence of scattering the conductance
takes on its universal value $K_{u}$.

To calculate the transmission coefficient $\mathcal{T}_{m}(\omega)$ in
Eq.~(\ref{conductance}) we use a Green function method similar to the one
previously reported in the scalar-wave calculation \cite{SC00}. The thermal
pathway is modelled as a rectangular elastic beam or waveguide of width $W$,
depth $d$, and length $L$ in the x-direction. We assume smooth top and bottom
surfaces corresponding to the epitaxial growth planes, and rough surfaces at
$y=\pm W/2+f_{\pm}(x)$ deriving from the lithographic processing. The
functions $f_{\pm}(x)$ define the roughness, which we assume to be independent
of the $z$ coordinate and small $f_{\pm}\ll W$.

Assuming isotropic elasticity theory, the displacement field $\mathbf{u}$
satisfies the wave equation:
\begin{equation}
\rho\partial^{2}u_{i}/\partial t^{2}-\partial T_{ij}/\partial x_{j}=0,
\label{wave}%
\end{equation}
where $\rho$ is the mass density, and $T_{ij}$ is the stress tensor,
$T_{ij}=C_{ijkl}\partial_{k}u_{l}$ with $C_{ijkl}$ the elastic modulus tensor.
Stress free conditions apply at the surfaces $T_{ij}n_{j}|_{S}=0$ with
$\mathbf{n}$ the surface normal. The Green function $G_{ij}(\vec{x},\vec
{x}^{\prime};t)$ is defined to satisfy the corresponding equation with a
source term $\delta_{ij}\delta(\mathbf{x}-\mathbf{x}^{\prime})\delta
(t-t^{\prime})$ on the right hand side, and stress free boundary conditions at
the smoothed surfaces $y=\pm W/2$. Using the completeness of the orthonormal,
smooth-surface, elastic modes $\mathbf{u}_{k}^{(n)}(\mathbf{x},\omega)$ at
frequency $\omega$ (with $k$ the wave number along the wave guide and $n$
labelling the transverse mode structure), the Fourier transformed Green
function can be written in the form
\begin{equation}
G_{ij}(\vec{x},\vec{x}^{\prime};\omega)=i\sum_{n}\,\,\frac{u_{i}^{(n)\ast
}(\vec{x}^{\prime},\omega)u_{j}^{(n)}(\vec{x},\omega)}{2\rho\omega v_{g}%
^{(n)}}, \label{Green-fcn}%
\end{equation}
The sum is over the elastic modes $n$ that propagate at frequency $\omega$:
the condition $\omega(k)=\omega$ with $\omega(k)$ the dispersion relation
defines a discrete set of wave numbers $k_{n}$ and corresponding mode indices
$n$. For monotonic $\omega(k)$ there is a single solution for $k$ on each
branch (defined by a continuous $\omega(k)$), but with anomalous dispersion
there may be multiple solutions. $v_{g}^{(n)}$ is the group velocity of mode
$n$ at frequency $\omega$.

The stress free boundary conditions for the propagating waves apply at the
actual (rough) surfaces; expanding in small $f_{\pm}$ lead to effective stress
boundary conditions at the smooth surfaces $y=\pm W/2$, which can then be
included using Green's theorem. These manipulations lead to the expression for
the scattered field $\mathbf{u}^{(sc)}$ to lowest order in the small roughness
in terms of the stress field of the incident wave at $y=\pm W/2$. The
back-scattered field from a unit-amplitude incident wave in mode $m$ is
\begin{equation}
\vec{u}^{(sc)}=\sum_{n,v_{g}^{(n)}<0}\frac{i\vec{u}^{(n)}(\vec{x},\omega
)}{2\rho\omega v_{g}^{(n)}}\int_{-\infty}^{\infty}f_{\pm}(x^{\prime}%
)\Gamma_{\pm}^{(m,n)}(x^{\prime},\omega)dx^{\prime},\label{scatter-field}%
\end{equation}
with
\begin{multline}
\Gamma_{\pm}^{(m,n)}=\int
dz\{\rho\omega^{2}u_{i}^{(m)}u_{i}^{(n)\ast
}-\label{Gamma}\\
E^{-1}[(T_{xx}^{(m)}T_{xx}^{(n)\ast}+T_{zz}^{(m)}T_{zz}^{(n)\ast}%
)-\sigma(T_{xx}^{(m)}T_{zz}^{(n)\ast}+T_{zz}^{(m)}T_{xx}^{(n)\ast})]+\mu
^{-1}T_{zx}^{(m)}T_{xz}^{(n)\ast}\}_{y=\pm W/2},
\end{multline}
(repeated indices $i$ and $\pm$ are to be summed over) where $E$
is Young's modulus, $\mu=E/2\left(  1+\sigma\right)  $ the shear
modulus, $\sigma$ the Poisson ratio, and $T_{ij}^{(n)}$ is the
stress field corresponding to the displacement field
$\mathbf{u}^{(n)}$. The mode sum in Eq.~(\ref{scatter-field}) is
confined to modes propagating to \emph{negative} $x$,
i.e.~$v_{g}^{(n)}(k_{n})<0$: usually these will be modes with
negative wave numbers $k_{n}$, but there are also regions of
anomalous dispersion which have a negative group velocity for
positive $k_{n}$. Notice that Eq.~(\ref{scatter-field}) is
comprised of separate kinetic, compression and
shear terms.%

\begin{figure}
[tbh]
\begin{center}
\includegraphics[
height=2.7717in,
width=3.6832in
]%
{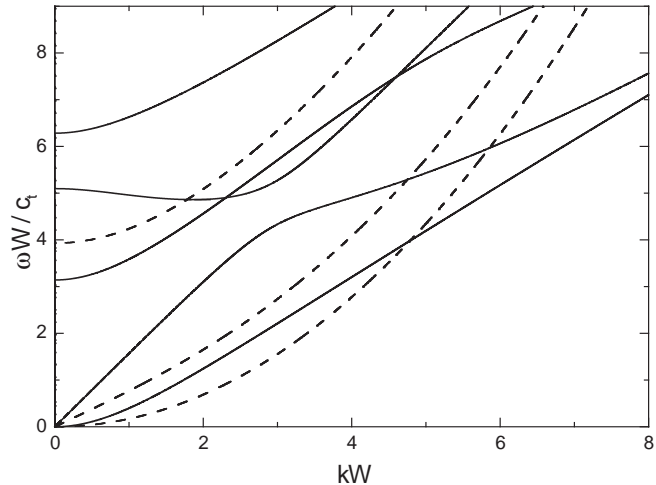}%
\caption{Dispersion relation for in-plane modes (solid) and flexural modes
(dashed) for a geometry ratio $d/W=0.375$ and Poisson ration $0.24$. The wave
numbers are scaled with the width $W$, and the frequencies by $W/c_{t}$ with
$c_{t}=\sqrt{\mu/\rho\text{.}}$}%
\label{dispersion}%
\end{center}
\end{figure}
Although the expression Eq.~(\ref{scatter-field}) applies quite generally to a
three dimensional waveguide with a rectangular cross-section \footnote{We have
specialized to the case of smooth top and bottom surfaces. A corresponding
development for rough surfaces on all the faces leads to a similar but lengthy
expression that will be presented elsewhere.}, there are no closed-form
expressions for the modes $\mathbf{u}^{(n)}$ in general, and so a direct
evaluation of the scattering would have to be done with numerical solutions
for the modes. Here, we instead use the \emph{thin plate approximation} $d\ll
W$ \cite{LF7,CL01}, which yields closed form expressions for the displacement
fields of the modes (in terms of a dispersion curve $\omega(k)$ given by
numerical solution of a simple transcendental equation). This approach
correctly accounts for the important properties of the elastic modes, for
example exactly reproducing the small $k$ dispersion relations including the
quadratic dispersion of the bend modes, and showing regions of anomalous
dispersion, whilst at the same time allowing analytic progress . Thin plate
theory should be quantitatively accurate for many mesoscopic experiments where
the thickness of the sample (formed by epitaxial growth) is often much less
than the width (given by lithography). The dispersion curves in the thin plate
approximation are shown in Fig.~(\ref{dispersion}). Note that the modes
polarized in the plane and perpendicular to the plane (flexural modes) do not couple.

An important simplification in thin plate theory is that all $z$-components of
the stress tensor $T_{iz},T_{zi}$ in Eq.~(\ref{Gamma}) may be put to zero
\cite{LF7}. The expression for the flux scattering rate (per unit length of
rough waveguide) from mode $m$ to mode $n$ at frequency $\omega$, defined as
$\left|  v_{g}^{(n)}/v_{g}^{(m)}\right|  $ times the ratio of the mode
intensities, simplifies to
\begin{equation}
\gamma_{m,n}(\omega)=\frac{\omega^{2}\tilde{g}(k_{m}-k_{n})}{2\left|
v_{g}^{(m)}v_{g}^{(n)}\right|  W^{2}}\left|  \int dz\left\{  \phi_{i}%
^{(m)}\phi_{i}^{(n)\ast}-\frac{Ek_{m}k_{n}}{\rho\omega^{2}}\phi_{x}^{(m)}%
\phi_{x}^{(n)\ast}\right\}  _{y=W/2}\right|  ^{2} \label{scatter-coeff}%
\end{equation}
where $\phi$ gives the transverse dependence of the modes $u_{i}%
^{(n)}=e^{ik_{n}x}\phi_{i}^{(n)}(y,z)$, and again back-scattering corresponds
to modes $n$ with negative group velocities which except for the regions of
anomalous dispersion correspond to negative $k_{n}$. To arrive at this
expression we have performed an ensemble average over the surface roughness
function $f_{\pm}(x)$, and $\tilde{g}(k)$ is the spectral density, i.e.~the
Fourier transform of the roughness correlation function $g(x)$ defined by%
\begin{equation}
\left\langle f(x)f(x^{\prime})\right\rangle =g(x-x^{\prime}),
\end{equation}
where $f$ is $f_{+}$ or $f_{-}$, and we assume independent scattering off the
two rough surfaces introducing a factor of $2$ into Eq.~(\ref{scatter-coeff}).
For weak scattering the transmission coefficient appearing in the equation for
the thermal conductance Eq.~(\ref{conductance}) can be obtained as
$\mathcal{T}_{m}=e^{-\gamma_{m}L}$ with $\gamma_{m}=\sum_{n,v_{g}^{(n)}%
<0}\gamma_{m,n}$, where we sum only over the back scattering, since scattering
into forward propagating modes does change the heat transport.

\begin{table*}[ptb]
\centering
\begin{tabular}
[c]{||c|c||}\hline\hline
in-plane & flexural\\\hline%
\begin{tabular}
[c]{c|c|c}%
cc & bb & bc,cb\\\hline
$2\bar{\omega}^{2}$ & $\sqrt{3}\bar{\omega}$ & $\frac{3^{5/4}}{2^{3/2}}%
\bar{\omega}^{3/2}$%
\end{tabular}
&
\begin{tabular}
[c]{c|c|c}%
tt & bb & tb,bt\\\hline
$\frac{9(1+\sigma)}{4}\left(  \frac{W\bar{\omega}}{d}\right)  ^{2}$ &
O$\left[  \left(  \frac{W\bar{\omega}}{d}\right)  ^{3}\right]  $ &
$\frac{3^{5/4}(1+\sigma)^{1/2}}{4}\left(  \frac{W\bar{\omega}}{d}\right)
^{3/2}$%
\end{tabular}
\\\hline\hline
\end{tabular}
\caption{Scattering coefficients for the zero onset frequency modes at low
frequencies: c denotes compression, b denotes bend, t denotes torsion, bb
denotes bend to bend scattering etc. Values are quoted for $\gamma_{m}W^{4}%
/\tilde{g}(0)$ as a function of scaled frequency
$\bar{\omega}=\omega c_{E}/W$. For the flexural bend to bend
scattering (bb) the terms in the braces in Eq.\
(\ref{scatter-coeff}) cancel to leading order resulting in very
small $O(\bar{\omega}^{3})$ scattering. There is no scattering
between
in-plane and flexural modes for the z-independent roughness assumed.}%
\label{Table1}%
\end{table*}

We first study the scattering behavior for unstructured ``white
noise'' roughness, $\tilde{g}(k)=\tilde{g}(0)$ independent of $k$.
At low frequencies only the four lowest modes with zero onset
frequency survive (the compression mode and a bending mode
polarized in the plane, and the torsion mode and a second bending
mode polarized perpendicular to the plane). Explicit expressions
for the scattering coefficients from these modes $\gamma
_{m}(\omega)W^{4}/\tilde{g}(0)$ are shown in Table \ref{Table1} as
a function of the scaled frequency $\bar{\omega}=\omega c_{E}/W$
with $c_{E}=\sqrt {E/\rho}$ the propagation speed of extension
waves in the beam. Notice the back-scattering from the compression
mode to the time reversed mode, and from the torsion mode to its
time reversed mode, have the $\omega^{2}$ dependence anticipated
in analogy with $1d$ Rayleigh scattering (as indeed was found in
the scalar-wave calculation). On the other hand scattering to or
from the bending modes introduces quite different power law
dependences, that will dominate at low frequencies. Most of this
novel behavior can be understood simply from the prefactor in
Eq.~(\ref{scatter-coeff}), $\gamma\varpropto
\omega^{2}/v_{g}^{\left(  m\right)  }v_{g}^{\left(  n\right)  }$,
since $v_{g}$ goes to a constant at small frequencies for the
compression and torsion modes, and is proportional to
$\omega^{1/2}$ for the bending modes. The remaining part of
Eq.~(\ref{scatter-coeff}),\emph{ }the integral over $z$, is $O(1)$
as $\omega\rightarrow0$, except for the flexural bending mode
back-scattering where the two contributions cancel at leading
order. Note also that the expressions for the flexural modes
involve additional factors of the width to depth ratio $W/d$, so
that in the thin plate limit these modes will be scattered more
strongly at a given $\omega$.

The low temperature thermal conductance can be derived directly from the low
frequency scattering expressions:\ if we write these as $\gamma L=A\bar
{\omega}^{p}$, then the corresponding contribution of the suppression of the
thermal conductance is
\begin{equation}
\delta K/K_{u}=AI_{p}(T/T_{E})^{p} \label{temp-power}%
\end{equation}
with $T_{E}=\hbar c_{E}/k_{B}W$ and $I_{p}$ the integral
\begin{equation}
I_{p}=\frac{3}{\pi^{2}}\int_{0}^{\infty}dy\frac{y^{p+2}e^{y}}{(e^{y}-1)^{2}}.
\end{equation}
giving the same power law for the temperature dependence.%

\begin{figure}
[tbh]
\begin{center}
\includegraphics[
height=2.7709in,
width=3.5405in
]%
{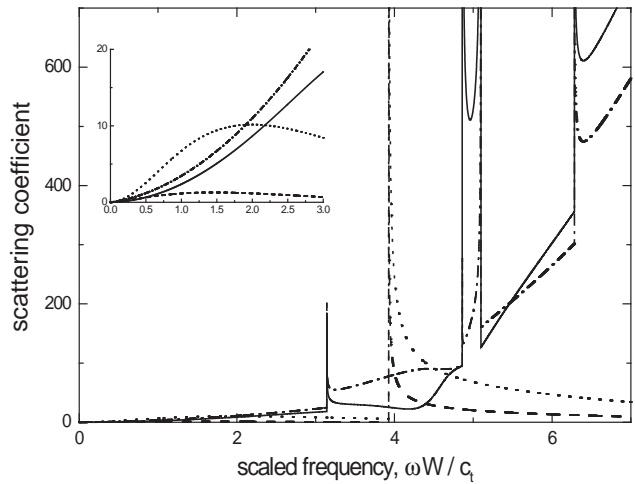}%
\caption{Scattering coefficient $\gamma_{m}W^{4}/\tilde{g}(0)$
of the lowest (zero onset frequency) modes to all
other modes for unstructured roughness: dashed-dotted -
compression; solid
- in-plane bend; dashed - flexural bend; dotted - torsion.}%
\label{individual-scattering}%
\end{center}
\end{figure}
Figure (\ref{individual-scattering}) shows the scattering coefficients over an
extended frequency range, again with $\tilde{g}(k)=\mathrm{const}$. The
phonons are strongly scattered at each mode onset frequency, as a result of
the zero group velocity, and the correspondingly large density of states to
scatter into. We find additional strong scattering in the regions of anomalous
dispersion (e.g.\ $\omega W/c_{t}\simeq5$), since again these portions of the
dispersion curve are quite flat. Also, since the group velocity of the
in-plane bending mode approaches zero, the scattering of this mode increases
relative to others as $\omega\rightarrow0$. This agrees with the low frequency
analysis, $\gamma\varpropto\omega$. (The back-scattering of the flexural
bending mode is reduced by the cancellation between the kinetic and stress
scattering.) For intermediate frequencies $0.08<\omega W/c_{t}<1.8$ the
scattering of the torsion mode is larger than for the other modes.%

\begin{figure}
[tbh]
\begin{center}
\includegraphics[
height=2.7709in,
width=3.3122in
]%
{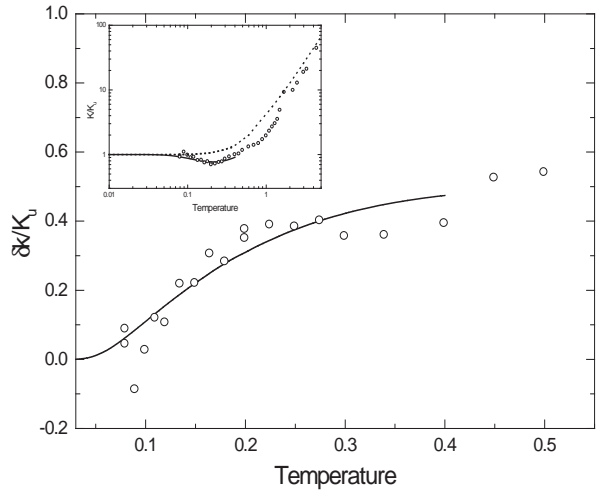}%
\caption{Thermal conductance change relative to the universal value $K_{u}$ as
a function of temperature: circles - data of Schwab et al.\
subtracted from the predicted ideal conductance; solid line -
predictions from the elasticity calculation. The insert shows the
thermal conductance relative to the universal value $K_{u}$ as a
function of temperature: circles - data of Schwab et al.; dotted
line - ideal conductance calculated using the xyz method; and
solid line - fit from elasticity theory with scattering. The
roughness parameters
used were $\delta/W=0.11$ and $k_{0}W=4.7$.}%
\label{bestfit}%
\end{center}
\end{figure}
With this detailed understanding, we now attempt to fit the experiments of
Schwab et al. We focus on the data below $0.4$K where thin plate theory
captures the relevant modes \footnote{We compare the $k\rightarrow0$ cut-off
frequencies, (which determine the ideal thermal conductance) obtained from
thin plate theory with those from numerical solutions of the full $3d$
elasticity equations using the xyz algorithm \cite{NAW97} for the thickness to
width ratio $d/W\simeq0.4$ appropriate to the experimental geometry.\ For
dimensionless frequencies $\omega W/c_{t}<8$ (with $c_{t}$ the propagation
speed of transverse waves), thin plate theory gives a good approximation.
Correspondingly, thin plate theory provides better than $5\%$ accuracy for the
ideal thermal conductance up to $0.4$K. The range of applicable temperature
will improve with decreasing $d/W$.}. To compare with experiment we first
subtract the \emph{measured} conductance from the ideal conductance predicted
using the cutoff frequencies \emph{calculated} numerically using the $xyz$
approach \cite{NAW97} using $d=60\mathrm{nm}$ and $W=160\mathrm{nm}$: this
gives us an estimate of the effect of scattering which we can compare with our
calculated scattering. The data processed in this way, Fig.~(\ref{bestfit}),
shows an abrupt onset of scattering at some non-zero temperature ($T\sim
0.08$K): this does not appear to be consistent with the power law behavior
predicted using unstructured roughness, Table \ref{Table1} and
Eq.~(\ref{temp-power}). Notice that the discrepancy is worse than anticipated
in the scalar-wave calculation \cite{SC00}, which gave larger exponents for
the power law dependence. The delay in the onset of scattering suggests that
the phonons with long wavelength are not much affected by the rough surfaces.
To accommodate this fact, we have investigated the scattering by roughness
centered around a finite length scale $k_{0}^{-1}$, using the parameterization%
\begin{equation}
\tilde{g}(k)=\sqrt{\pi}\delta^{2}ae^{-a^{2}(k-k_{0})^{2}}%
\end{equation}
For $k_{0}a\gtrsim1$ this results in strongly reduced scattering at long
wavelengths. There are three roughness parameters that now characterize the
surfaces: the roughness amplitude $\delta$, the correlation length $a$, and
shift $k_{0}$. We obtain a reasonable fit to the low temperature onset of the
scattering using parameters $a/W=3$, $\delta/W=0.11$, and $k_{0}W=4.7$. These
parameters correspond to significant roughness at length scales comparable to
the width of the device, which appears consistent with the electron-micrograph
of the structure used \cite{Spc}.

In summary, we have examined the scattering of phonons by surface roughness
and the effect on the universal thermal conductance. At low temperature, the
scattering shows strong dependence on mode structure. In this temperature
range, the elasticity model provides a better understanding of the scattering
behavior and a reasonable fit to the experimental data if we assume that the
roughness is concentrated at a length scale comparable to the lithographic
scale. Further experiments at lower temperatures would provide a useful test
of the predictions of the theory.

\begin{acknowledgments}
This work was supported by NSF grant no. DMR-9873573.
\end{acknowledgments}

\end{document}